# Cryptanalysis of a New Knapsack Type Public-Key Cryptosystem

Roohollah Rastaghi

*Department of Electrical Engineering, Aeronautical University of Since & Technology, Tehran, Iran*

r.rastaghi59@gmail.com

*Abstract*— Recently, Hwang et al. introduced a knapsack type public-key cryptosystem. They proposed a new algorithm called permutation combination algorithm. By exploiting this algorithm, they attempt to increase the density of knapsack to avoid the low-density attack.

We show that this cryptosystem is not secure, as it based on basic Merkel-Hellman knapsack cryptosystem and because of the superincreasing structure, we can use shamir's attack on the basic Merkel-Hellman knapsack to break this cryptosystem.

## I. Introduction

IN 1976, Diffie and Hellman [3] introduced the notion of the public-key cryptography. Until that time, most public-key cryptosystems (PKC) fall into one of the two below categories [1]:

- Public-key cryptosystems based on hard number-theoretic problems: e.g., RSA [13], ElGamal [4] and ….
- Public-key cryptosystems based on subset sum or subset product problems: e.g., Merkle-Hellman [9], Chor-Rivest [1], Morri-Kasahara [11], Naccache-Stern [12],… .

Unlike hard number-theoretic problems, the knapsack problem has been proven to be *NP-complete* [10]. That is, there is no polynomial algorithm will be invented to solve the knapsack problem.

Since its Merkle-Hellman proposal, knapsack PKCs had been widely studied, and many knapsack-based PKCs were developed. There is no question that knapsack PKCs still warrant continuous researches, as a result of the NP-completeness nature, the faster speed and a desire to have a wide variety of available cryptosystems. Nowadays, researchers reconsider knapsack public-key cryptography also because Shor [15] showed that integer factorization and discrete logarithm problems can be easily solved by using quantum computers. Therefore, traditional PKC schemes based on the two problems cannot be used to provide privacy protections any longer, and PKC schemes secure in quantum computing environment are needed to be developed. Although the underlying problem is NP-complete, but almost all knapsack cryptosystems were shown insecure in that they are vulnerable to some known attacks such as: low density attack [2,6], Shamir's attack [14] and diophantine approximation attack [17]. This vulnerability is due to the special structure of the private key and the mathematical methods that public key (public knapsack) was built from the private key.

In this paper, we analyze security of the Hwang *et al*. cryptosystem [5]. We show that due to similarity of the key generation algorithm of their scheme with the basic Merkel-Hellman cryptosystem, we can use Shamir's attack to obtain equivalent private keys.

The rest of this paper is organized as follows. In the next section, we briefly explain subset sum problem and the basic Merkle-Hellman cryptosystem. Then, in Section 3, we review the Shamir's attack. Hwang *et al.'s* knapsack cryptosystem will be presented in Section 4 and cryptanalysis of this system will be given in Section 5.

## II. The Subset Sum Problem And The Basic Merkle-Hellman Cryptosystem

The subset sum problem is stated as follows: given a set of positive integers $(a_1, \ldots, a_n)$ and positive integer $s$. Whether there is a subset of the $a_i$s that sums to $s$. This is equivalent to determine whether there are variables $(x_1, \ldots, x_n)$ such that

$$s = \sum_{i=1}^{n} a_i x_i, \qquad x_i \epsilon \{0,1\}, \qquad 1 \leq i \leq n.$$

If the set of positive integers $(a_1, \ldots, a_n)$ be a superincreasing sequence, e.g. $b_i > \sum_{j=1}^{i-1} b_j$, $i \geq 2$, then the knapsack problem is solvable in polynomial time.

The basic Merkel-Hellman knapsack cryptosystem uses a superincreasing sequence as a private key. This cryptosystem is as follows:

**Key generation.** The designer chooses a superincreasing sequence $(b_1, \ldots, b_n)$ and two large positive integers $w$ and $p$, such that

$$p > \sum_{i=1}^{n} b_i \quad , \quad \gcd(w,p) = 1 .$$

He also selects a permutation $\pi$ of $\{1, 2, \ldots, n\}$ and then transforms the easily solved knapsack $B$ into trapdoor knapsack $(a_1, \ldots, a_n)$ via the relation

$$a_i = w \cdot b_{\pi(i)} \mod p. \quad (1)$$

The public key is $(a_1, \ldots, a_n)$ and the private key is $\{(b_1, b_2, \ldots, b_n), W, P, \pi\}$.

**Encryption.** To encrypt message $m = (m_1, \ldots, m_n)$, he computes

$$c = \sum_{i=1}^{n} a_i m_i,$$

and sends it to the receiver.

**Decryption:** To recover plaintext $m$ from ciphertext $c$, the receiver should perform the following steps.
1) Compute

$$d = cw^{-1} \mod p.$$

2) With his private key, $(b_1, \ldots, b_n)$, solve a superincreasing subset sum problem and find integers $(r_1, \ldots, r_n)$, $r_i \in \{0, 1\}$ such that

$$d = \sum_{i=1}^{n} d_i r_i \mod p.$$

Note that since $p > \sum_{i=1}^{n} b_i$ hence $d = \sum_{i=1}^{n} d_i r_i$.
3) The message bits are

$$m_i = r_{\pi(i)}, \quad i = 1, 2, \ldots, n.$$

### III. SHAMIR ATTACK ON BASIC MERKLE-HELLMAN KNAPSACK CRYPTOSYSTEM

In 1982, Adi Shamir [14] shows that modular multiplication cannot perfectly hide the superincreasing sequence (private key), and hence, all the equation of the form

$$c = \sum_{i=1}^{n} x_i a_i, \quad x_i \in \{0,1\},$$

can be solved in polynomial time. This approach originates with Shamir [14] although we follow the presentation of Lagarias [7].

Such as Hwang *et al.'s* knapsack cryptosystem [5], we assume that no permutation is used. Hence equation (1) can be written as follows:

$$a_i = w \cdot b_i \mod p.$$

Let $U = w^{-1} \mod p$ where $1 \leq U < p$. We have

$$b_i = Ua_i \mod p.$$

This means that for $1 \leq i \leq n$, there exists some integers $k_i$ such that

$$a_i U - k_i p = b_i$$

and $0 \leq k_i < a_i$. Hence,

$$0 \leq U/p - k_i/a_i = b_i/a_i p. \quad (2)$$

Since the $b_i$s are superincreasing we have $b_i < p/2^{n-i}$ and so

$$0 \leq U/p - k_i/a_i < 1/a_i 2^{n-i}.$$

In particular, the right side of $U/p - k_1/a_1 < 1/(a_1 2^{n-1})$ is very small. Hence we can assume $U/p \approx k_1/a_1$.

We now observe that to break the basic Merkle-Hellman knapsack it is sufficient to find any pair $(U', p')$ of positive integers such that $U'a_i \mod p'$ is a superincreasing sequence (or similar enough to a superincreasing sequence that one can solve the subset sum problem). We show that if $k_1/a_1$ is close enough to $U/p$, then $(U', p') = (k_1, a_1)$.

Subtracting the case $i = 1$ of equation (2) from the $i$-th gives

$$\frac{k_1}{a_1} - \frac{k_i}{a_i} = \frac{b_i}{a_i p} - \frac{b_1}{a_1 p} = \frac{a_1 b_i - a_i b_1}{a_1 a_i p}$$

and so, for $2 \leq i \leq n$,

$$|a_i k_1 - a_1 k_i| = \frac{|a_1 b_i - a_i b_1|}{p} < \frac{2pb_i}{p} = 2b_i < \frac{p}{2^{n-i-1}}. \quad (3)$$

Taking $p' = a_1$ and $U' = k_1$ then $U'a_i \mod p'$ is very close to a superincreasing sequence.

Since $a_1$ is public, It remains to compute the integer $k_1$ such that equation (3) holds, given only the integers $a_1, \ldots, a_n$. Another way to write equation (3) is

$$\left| \frac{a_i}{a_1} - \frac{k_i}{k_1} \right| = \frac{p}{a_1 k_1 2^{n-i-1}},$$

and one sees that the problem is precisely simultaneous diophantine approximation. We can use lattice based reduction algorithm for solving simultaneous diophantine approximation. Performing lattice basis reduction one obtains a guess for $k_1$. We now set $U' = k_1$ and $p' = a_1$ and computes $U'a_i \mod p'$ for $2 \leq i \leq n$. This is a superincreasing sequence. We then compute $U'c \mod p'$ for any challenge ciphertext $c$ that is decrypted using the superincreasing sequence, and therefore message is recovered.

### IV. DESCRIPTION OF HWANG ET AL.'S CRYPTOSYSTEM

Hwang's cryptosystem is based on the Merkle-Hellman cryptosystem. In the key generation stage, each user chooses a superincreasing sequence $B = \{b_1, \ldots, b_{1360}\}$ as secret key.

i.e.
$$b_i > \sum_{j=1}^{i-1} b_j \quad (i = 1, 2, \ldots, 1360).$$

$W$ and $W'$ are secret modular multipliers such that

$$\gcd(P, W) = 1, \quad P > \sum_{i=1}^{1360} b_i \quad \text{and} \quad W \times W' = 1 \bmod P.$$

Each user transfers superincreasing sequence $B = \{b_1, \ldots, b_{1360}\}$ into a pseudorandom sequence $A = \{a_1, \ldots, a_{1360}\}$ as follows:

$$a_i = b_i . W \bmod P, \quad (1 \leq i \leq 1360). \quad (4)$$

Further, each user chooses a random 170× 256 binary matrix H, a vector $R = (r_1, \ldots, r_{256})^T$ and a vector $HR = (hr_1, \ldots, hr_{170})^T$ to satisfy the following equation:

$$H.R = HR \bmod n$$

$$\begin{pmatrix} h_{1,1} & \cdots & h_{1,256} \\ \vdots & \ddots & \vdots \\ h_{170,1} & \cdots & h_{170,256} \end{pmatrix} \cdot \begin{pmatrix} r_1 \\ \vdots \\ r_{256} \end{pmatrix} = \begin{pmatrix} hr_1 \\ hr_2 \\ \vdots \\ hr_{170} \end{pmatrix} \bmod n$$

$$= \begin{pmatrix} 2^0 \\ 2^1 \\ \vdots \\ 2^{169} \end{pmatrix} \bmod n$$

$$hr_i = 2^{i-1} = \sum_{j=1}^{256} h_{i,j} r_j \bmod n \quad (i = 1, 2, \ldots, 170).$$

Let $H(.)$ be a one-way hash function.
The public key is $(A, R)$ and the private key is $(H, B, W, W', P)$.

They present a permutation algorithm and want this algorithm to ensure the security of the cryptosystem. The permutation algorithm is as follows:

1) Define an original sequence

$$D_0 = \{E_n, E_{n-1}, E_{n-2}, \ldots, E_5, E_4, E_3, E_2, E_1\}.$$

2) Recombine all the elements of the original sequence $D_0$ which obtain $(n! - 1)$ sequences $D_1, \ldots, D_{(n!-1)}$. The sequences $D_i(i = 1, 2, \ldots, n! - 1)$ are defined as follows:

$$D_0 = \{E_n, E_{n-1}, E_{n-2}, \ldots, E_5, E_4, E_3, E_2, E_1\}$$
$$D_1 = \{E_n, E_{n-1}, E_{n-2}, \ldots, E_5, E_4, E_3, E_1, E_2\}$$
$$D_2 = \{E_n, E_{n-1}, E_{n-2}, \ldots, E_5, E_4, E_2, E_3, E_1\}$$
$$D_3 = \{E_n, E_{n-1}, E_{n-2}, \ldots, E_5, E_4, E_2, E_1, E_3\}$$
$$D_4 = \{E_n, E_{n-1}, E_{n-2}, \ldots, E_5, E_4, E_1, E_3, E_2\}$$
$$D_5 = \{E_n, E_{n-1}, E_{n-2}, \ldots, E_5, E_4, E_1, E_2, E_3\}$$
$$\vdots$$

$$D_{n!-1} = \{E_1, E_2, E_3 E_4, \ldots, E_{n-2}, E_{n-1}, E_n\}$$

3) Suppose we can compute $D_m$ for $1 \leq m \leq n! - 1$. $m$ can be written as

$$m = \sum_{i=1}^{n} F_i (n-i)!, \quad 0 \leq F_i \leq n - i$$

each sequence has an own corresponding value called the factorial carry value $\{F_n, F_{n-1}, \ldots, F_2, F_1\}$. Using the factorial carry value, we can efficiently obtain any sequence. Let $m = 6$ and we want determine the sequence $D_6$. We can write

$$6 = 0 \times (n-1)! + \cdots + 1 \times 3! + 0 \times 2! + 0 \times 1! + 0$$

So the factorials carry value of $D_6$ is:
$$\{F_n, F_{n-1}, \ldots, F_2, F_1\} = \{0, 0, \ldots, 0, 1, 0, 0, 0\}$$

4) With the knowledge of the original sequence $\{E_n, E_{n-1}, E_{n-2}, \ldots, E_5, E_4, E_3, E_2, E_1\}$ and the factorial carry value $\{0,0,0,\ldots,0,1,0,0,0\}$ of $D_6$, we can compute sequence $D_6$ as follows:

Get $E_n$ by introducing $F_n = 0$. Here, the remaining elements in the sequence are $\{E_{n-1}, E_{n-2}, \ldots, E_5, E_4, E_3, E_2, E_1\}$.
Get $E_{n-1}$ by introducing $F_{n-1} = 0$. Here, the remaining elements in the sequence are $\{E_{n-2}, \ldots, E_5, E_4, E_3, E_2, E_1\}$.
$\vdots$
Get $E_5$ by introducing $F_5 = 0$. Here, the remaining elements in the sequence are $\{E_4, E_3, E_2, E_1\}$.
Get $E_3$ by introducing $F_4 = 1$. Here, the remaining elements in the sequence are $\{E_4, E_2, E_1\}$.
Get $E_4$ by introducing $F_3 = 0$. Here, the remaining elements in the sequence are $\{E_2, E_1\}$.
Get $E_2$ by introducing $F_2 = 0$. Here, the remaining elements in the sequence are $\{E_1\}$.
Get $E_1$ by introducing $F_1 = 0$.
Therefore, the sequence $D_6$ is:

$$\{E_n, E_{n-1}, E_{n-2}, \ldots, E_5, E_3, E_4, E_2, E_1\}.$$

**Encryption:** The sender A executes the following steps to generate the ciphertext $C$ of the message $M$.

1) Compute the digest $D$ of $M$ as
$$D = H_{1024}(M).$$
2) Compute $D' = D \bmod 170!$
3) Compute the factorial carry value $U = \{u_1, u_2, \ldots, u_{170}\}$ of $D'$ where

$$D' = u_1 \times 169! + u_2 \times 168! + \cdots + u_{170} \times 0!$$

4) Divide B's public key vector $A_b = \{a_{b1}, a_{b2}, \ldots, a_{b1360}\}$ into 8 subset public key vectors. Each subset public key vector has 170 elements.

$$A_b = \{(a_{b1}, a_{b2}, \ldots, a_{b170}),$$

$$(a_{b171}, a_{b172}, \ldots, a_{b340}),$$
$$\vdots$$
$$(a_{b1191}, a_{b1192}, \ldots, a_{b1360})\}.$$

5) Recombine each subset public key vector using $U = \{u_1, u_2, \ldots, u_{170}\}$ by means of the Permutation Combination Algorithm. $A$ chooses each subset public key vector in the first 128 elements. Then, $A$ will obtain 1024 elements $A_{bu} = \{au_{b1}, au_{b2}, \ldots, au_{b1024}\}$.

6) $M$ is divided into $\{M_1, M_2, \ldots, M_j\}$. Each $M_k$ ($k = 1, 2, \ldots, j$) is a 1024-bit message.

$$M_k = \{x_{k,1}, \ldots, x_{k,1024}\}$$

7) The corresponding ciphertext $C_k$ is given as the product of $A_{bu}$ and $M_k$ ($k = 1, 2, \ldots, j$).

$$C_k = \sum_{i=1}^{1024} au_{bi} \times x_{k,i}$$

The ciphertext is $C = \{C_1, \ldots, C_j\}$. $A$ sends $(C, D')$ to $B$ through the insecure channel.

**Decryption:** After receiving $C$ and $D'$, $B$ executes the following steps to derive $M$ from $(C, D')$:

1) Compute the factorial carry value $U = \{u_1, u_2, \ldots, u_{170}\}$ of $D'$ where

$$D' = u_1 \times 169! + u_2 \times 168! + \cdots, u_{170} \times 0$$

2) Divide his secret key vector $B_b = \{b_{b1}, b_{b2}, \ldots, b_{b1360}\}$ into 8 subset public key vectors.

$$B_b = \{(b_{b1}, b_{b2}, \ldots, b_{b170}),$$
$$(b_{b171}, b_{b172}, \ldots, b_{b340}),$$
$$(b_{b341}, b_{b342}, \ldots, b_{b510}),$$
$$(b_{b511}, b_{b512}, \ldots, b_{b680}),$$
$$\vdots$$
$$(b_{b1191}, b_{b1192}, \ldots, b_{b1360})\}.$$

3) Recombine each subset public key vector using $U = \{u_1, u_2, \ldots, u_{170}\}$ by means of the Permutation Combination Algorithm. $B$ chooses each subset public key vector in the first 128 elements. Then, $B$ will obtain 1024 elements $B_{bu} = \{bu_{b1}, bu_{b2}, \ldots, bu_{b1024}\}$. However $B_{bu} = \{bu_{b1}, bu_{b2}, \ldots, bu_{b1024}\}$ is still a superincreasing sequence.

4) Divide $C$ into $C = \{C_1, \ldots, C_j\}$. Each $C_k$ ($k = 1, 2, \ldots, j$) is a 1024-bit ciphertext.

5) Compute the recombine message $Mre_k$, which is given as the product of $C_k$ and $W'$ ($k = 1, 2, \ldots, j$).

$$Mre_k = C_k \times W' \bmod P$$
$$= \sum_{i=1}^{1024}(au_{bi} \times x_{k,i}) \times W' \bmod P$$
$$= \sum_{i=1}^{1024}(bu_{bi} \times W \times x_{k,i}) \times W' \bmod P$$
$$= \sum_{i=1}^{1024} bu_{bi} \times x_{k,i} \bmod P$$

So the receiver solves this superincreasing knapsack problem and then obtains the message $M$.

Obviously, there is no difference between the cryptosystem above and the original Merkle-Hellman except the omission of the permutation $\pi$ in the key generation stage.

## V. ATTACKING THE CRYPTOSYSTEM

As we mentioned in section IV, Hwang *et al.* suppose that superincreasing sequence $B = \{b_1, \ldots, b_{1360}\}$ is private key and produce public key from equation (4):

$$a_i = b_i.W \bmod P, \quad 1 \le i \le 1360.$$

Let $U = W^{-1} \bmod P$. We have

$$b_i = a_i.U \bmod P, \quad 1 \le i \le 1360$$

We can solve simultaneous diophantine approximation, as described in section III, and find a pair of integers $(U', P')$ such that $U'/P'$ is close to $U/P$. With this pairs, we can now compute integers

$$b'_i = a_i.U' \bmod P', \quad 1 \le i \le 1360$$

which form a superincreasing sequence. This sequence can then be used in place of to secret key vector $B = (b_1, \ldots, b_{1360})$.

On the other hand, we can eavesdrop pair $(D', C)$ from insecure channel and hence we can compute factorial carry value $U = \{u_1, u_2, \ldots, u_{170}\}$ of $D'$ where

$$D' = u_1 \times 169! + u_2 \times 168! + \cdots + u_{170} \times 0!$$

We divide recovered pseudo secret key vector $B' = \{b'_1, \ldots, b'_{1360}\}$ into 8 subset public key vectors:

$$B' = \{(b'_1, b'_2, \ldots, b'_{170}),$$
$$(b'_{171}, b'_2, \ldots, b'_{340}),$$
$$\vdots$$
$$(b'_{1191}, b'_2, \ldots, b'_{1360})\}$$

and recombine each subset public key vector using $U = \{u_1, u_2, \ldots, u_{170}\}$ by means of the permutation combination algorithm. We choose each subset public key vector in the first 128 elements. Then, we will obtain 1024 elements $B'_u = \{b'u_1, b'u_2, \ldots, b'u_{1024}\}$.

Divide $C$ into $C = \{C_1, \ldots, C_j\}$ and with computed pair $(U', P')$ we can compute:

$$Mre_k = C_k \times U' \bmod P'$$
$$= \sum_{i=1}^{1024}(au_i \times x_{k,i})U' \bmod P'$$
$$= \sum_{i=1}^{1024}(b'u_i \times x_{k,i}) \bmod P'$$

Note that $au_i \times U' \bmod P' = b'u_i$.

Since $\{b'u_1, b'u_2, \ldots, b'u_{1024}\}$ is a superincreasing sequence, we can solve this easy knapsack problem for $1 \leq k \leq j$ and therefore we can obtain the message $M = \{M_1, M_2, \ldots, M_j\}$.

## VI. Conclusion

We considered a new knapsack-based PKC scheme. This scheme uses a permutation algorithm in the encryption phase to avoid the low density attack by keeping the density high. As we showed, this scheme is vulnerable, since like original Merkle-Hellman cryptosystem uses a superincreasing sequence as private key and attempt to hide this sequence with modular multiplication. But as Shamir showed, the modular multiplication cannot perfectly hide the superincreasing sequence. To avoid this attack, we can choose another easy knapsack problem, for example, like this presented in [18] or we do not use modular multiplication to produce the public key.

## References

[1] B. Chor and R.L. Rivest, "A knapsack-type public key cryptosystem based on arithmetic in finite fields", *IEEE Trans. Inform. Theory*, vol. 34, pp. 901–909, September 1988.

[2] M. J. Coster, B. A. LaMacchia, A. M. Odlyzko, and C. P. Schnorr, "An improved low-density subset sum algorithm, in Advances in Cryptology", *EUROCRYPT'91, Lecture Notes in Computer Science*, vol. 547, pp. 54–67, 1991.

[3] W. Diffiee, M. Hellman, "New Directions in Cryptography", *IEEE Transaction on Information Theory,* IT-22 (6), pp. 644-654, 1976.

[4] T. ElGamal, "A public-key cryptosystem and a signature scheme based on discrete logarithms", *IEEE Transactions on Information Theory*, vol. IT-31, pp. 469–472, July 1985.

[5] M. S. Hwang, C. C. Lee, and S. F. Tzeng, "A New Knapsack Public-Key Cryptosystem Based on Permutation Combination Algorithm", *International Journal of Applied Mathematics and Computer Sciences* vol. 5; 1, pp. 33-38, Winter 2009.

[6] J. C. Lagarias and A. M. Odlyzko, "Solving low-density subset sum problems", *J. Ass. Comput. Much.* vol. 32, no. 1, pp. 229-246. Jan. 1985.

[7] J. C. Lagarias, "Performance Analysis of Shamir's Attack on the Basic Merkle-Hellman Knapsack Public Key Cryptosystem", *Proc. 11th Intern. Colloquium on Automata, Languages and Programming (ICALP)*, Lecture Notes in Computer Science, vol. 172, pp. 312-323, Springer-Verlag, Berlin, 1984.

[8] K. Lenstra, H. W. Lenstra Jr., and L. Lovász, "Factoring Polynomials with rational coefficients", *Muth. Ann.,* vol.261, pp. 515-534, 1982.

[9] R. Merkle and M. E. Hellman, "Hiding information and signatures in trapdoor knapsack," *IEEE Trans. Inform. Theory*, vol. IT-24, pp. 525–530, Sept 1978.

[10] R. Michael, and S. David, *Computers and Intractability: A guide to the theory of NP-completeness*. W. H. Freeman & Co., San Francisco, 1979.

[11] M. Morii, M. Kasahara, "New public key cryptosystem using discrete logarithm over GF(p)," *IEICE Transactions*, vol. J71-D, pp. 448-453, 1988

[12] D. Naccache and J. Stern, "A new public-key cryptosystem," *Advances in Cryptology, euro crypt '97, Lecture Notes in computer Science*, vol. 1233, pp 27-36. Springer-Verlag, 1997.

[13] R. L. Rivest, A. Shamir, and L. Adleman, "A method for obtaining digital signatures and public key Cryptosystems," *Communications of the ACM*, vol. 21, pp. 120–126, Feb. 1978.

[14] A. Shamir, "A Polynomial-time Algorithm for Breaking the Basic Merkle-Hellman Cryptosystem," *Proceedings of the IEEE Symposium on Foundations of Computer Science, New York*, pp. 145-152, 1982.

[15] P.W. Shor, "Polynomial-time algorithms for prime factorization and discrete logarithms on a quantum computer," *SIAM Journal of Computing*, vol. 26, pp. 1484-1509, 1997.

[16] S. Vaudenay, "Cryptanalysis of the Chor-Rivest cryptosystem," *Advances in Cryptology CRYPTO 98, Lecture Notes in computer Science,* vol. 1462, pp. 243-256, Springer- Verlag, Berlin, 1998.

[17] B. Wang and Y. Hu, "Diophantine approximation attack on a fast public key cryptosystem," *ISPEC 2006, Lecture Notes in computer Science,* vol.3903, pp. 25-32, Springer-Verlag, Berlin, 2006.

[18] W. Zhang, B. Wang and Y. Hu, "A New Knapsack Public-Key Cryptosystem," *IEEE, Fifth International Conference on Information Assurance and Security*, pp.53-56, 2009.